\documentclass[12pt]{article}
\usepackage{hyperref}
\usepackage{amsmath}
\usepackage{graphicx}
\usepackage{amssymb} 
 \usepackage{xcolor}
\definecolor{myblue}{rgb}{0.14,0.11,0.49}
\definecolor{myred}{rgb}{0.74,0.22,0.15}
\definecolor{mygreen}{rgb}{0.05,0.52,0.42}
\definecolor{myyellow}{rgb}{0.96,0.92,0.13}
\definecolor{myorange}{rgb}{1,0.61,0.36}
\definecolor{mypurple}{rgb}{0.71,0.02,1}
\definecolor{noir}{gray}{0.} 

\newcommand{\Mat}[1]{{{\boldsymbol{#1}}}}
\newcommand{\abs}[1]{\left\vert#1\right\vert}
\def\be{\begin{equation}}
\def\ee{\end{equation}}
\def\bea{\begin{eqnarray}}
\def\eea{\end{eqnarray}}
\def\bi{\begin{itemize}}
\def\ei{\end{itemize}}
\def\noi{\noindent}
\def\dd{\mathrm{d}}

\date{}

\title{On the non-uniqueness problem of the covariant Dirac theory and the spin-rotation coupling}

\author{Mayeul Arminjon\\
\small\it Laboratory ``Soils, Solids, Structures, Risks'', 3SR\\ \small\it (CNRS and Universit\'es de Grenoble: UJF, Grenoble-INP),\\\small\it BP 53, F-38041 Grenoble cedex 9, France.}

\begin{document}
\maketitle

\begin{abstract} 
\noi Gorbatenko \& Neznamov [arXiv:1301.7599] recently claimed the absence of the title problem. In this paper, the reason for that problem is reexplained by using the notions of a unitary transformation and of the mean value of an operator, invoked by them. Their arguments actually aim at proving the uniqueness of a particular prescription for solving this problem. But that prescription is again shown non-unique. Two Hamiltonians in the same reference frame in a Minkowski spacetime, only one of them including the spin-rotation coupling term, are proved to be physically non-equivalent. This confirms that the reality of that coupling should be checked experimentally.\\

\noi {\bf Keywords:} Dirac Hamiltonian; curved spacetime; unitary transformation; rotating frame\\

\end{abstract}
\newpage

\section{Introduction}\label{Intro}

\noi In a recent preprint \cite{GorbatenkoNeznamov2013}, Gorbatenko \& Neznamov write that ``publications have emerged again \cite{A43, A47, A49}, which declare and provide grounds for the assertion that the Dirac theory is non-unique in a curved and even flat spacetime". They announce that, in contrast: ``in this work we again assert the absence of the non-uniqueness problem of the Dirac theory in a curved and flat spacetime and illustrate this with a number of examples." The aim of this paper is to show that they do not prove a such thing and that their examples do not and can not do that either. Their arguments do not address the former proof \cite{A43} of the generic non-uniqueness of the Hamiltonian and energy operators associated in a given reference frame with the (generally-)covariant Dirac equation --- be it in a curved or in a flat spacetime, indeed. Nor do their arguments answer my former proof \cite{A47} that their algorithm based on going to a special kind of tetrad leaves the Hamiltonian and energy operators ambiguous. That proof used a counterexample that is relevant precisely to my discussion of the spin-rotation coupling \cite{A49}, commented on in their Examples 6 and 7.\\

\section{Physically equivalent operators}\label{Equivalent Operators}

I fully agree with Gorbatenko \& Neznamov that the mere ``demonstration that the form of Dirac Hamiltonians depends on the choice of tetrads" would be ``absolutely insufficient" to ``demonstrate the non-equivalence of Dirac Hamiltonians". However, in spite of what they state, this was not at all the approach followed in the paper in which, together with F. Reifler, we proved the non-uniqueness of the Hamiltonian and energy operators of the covariant Dirac theory \cite{A43}. On the contrary, in the paper \cite{A43}, we began the study of that problem with carefully establishing the condition under which one may say that two versions of a quantum-mechanical operator such as the Hamiltonian, got by choosing two admissible coefficient fields in the covariant Dirac equation, are {\it physically equivalent.} Indeed the non-uniqueness problem is related, though not in a trivial way, with the fact that there is a vast continuum of different choices for the coefficient fields of the covariant Dirac equation. Any two such fields, thus any two fields of Dirac matrices $\gamma ^\mu $ and $\widetilde{\gamma }^\mu $, are related together by a ``local" similarity transformation, given by a non-singular complex matrix $S(X)$ that depends smoothly on the point $X$ in the spacetime $\mathrm{V}$:
\footnote{\label{A constant}\
In this paper we shall consider only the standard version of the covariant Dirac equation, or ``Dirac-Fock-Weyl" equation (DFW equation for short). For DFW, the hermitizing matrix $A$ is a constant matrix which is invariant under any admissible local similarity transformation $S$, i.e., any one got from ``lifting" a local Lorentz transformation applied to the (orthonormal) tetrad field \{Ref. \cite{A42}, Eq. (104) and below\}. Then the coefficient fields are indeed reduced to the field $\gamma ^\mu \, (\mu =0,...,3)$. In standard practice (including in Ref. \cite{GorbatenkoNeznamov2013}), one has even $A=\gamma ^{\natural 0}$, where $(\gamma ^{\natural \alpha })$ is some special set of Dirac matrices for the Minkowski spacetime in Cartesian coordinates. 
}
\be \label{similarity-gamma}
\widetilde{\gamma} ^\mu =  S^{-1}\gamma ^\mu S, \quad \mu =0,...,3.
\ee

\vspace{3mm}
We noted first that, with each of the two different coefficient fields: $\gamma ^\mu$ and $\widetilde {\gamma}^\mu$, corresponds a unique Hilbert scalar product. Explicitly:
\be \label{Hermitian-sigma=1-g}
(\Psi  \mid \Phi  ) \equiv \int \Psi^\dagger \sqrt{-g}\,A\gamma ^0\, \Phi\ \dd^ 3{\bf x}
\ee
for the first field, $\gamma ^\mu $, and 
\be \label{Hermitian-tilde}
(\Xi  \,\widetilde { \mid} \,\Omega   ) \equiv \int \Xi ^\dagger \sqrt{-g}\,\widetilde {A}\,\widetilde {\gamma}^0 \,\Omega \,\dd^ 3{\bf x},\qquad \widetilde {A}\equiv S^\dagger AS
\ee
for the second one, $\widetilde {\gamma}^\mu$ (with $S^\dagger AS=A$ if $S$ is an admissible similarity transformation for DFW). Thus, with the first coefficient field $\gamma ^\mu $, the wave function $\Psi $ lives in a Hilbert space $\mathcal{H}$ and, with the second coefficient field $\widetilde {\gamma}^\mu$, the wave function $\Xi  $ lives in a different Hilbert space $\widetilde{\mathcal{H}}$. Moreover, when applied precisely to the {\it wave function,} the similarity transformation $S$ defines a transformation $\mathcal{U}$ from the first Hilbert space $\mathcal{H}$ onto the second one,  $\widetilde{\mathcal{H}}$:
\be\label{psitilde=S^-1 psi} 
\mathcal{U}\Psi \equiv \widetilde{\Psi}\equiv S^{-1}\Psi,\quad i.e., \ (\mathcal{U}\Psi)(X) \equiv S(X)^{-1}\Psi(X).
\ee
This one-to-one mapping is a (linear) isometry \cite{A43}, or in other words a {\it unitary transformation $\mathcal{U}$} of $\mathcal{H}$ onto $\widetilde{\mathcal{H}}$, for we have from Eqs. (\ref{similarity-gamma}) to (\ref{psitilde=S^-1 psi}):
\be \label{tilde=isometry}
\forall \Psi ,\Phi \in \mathcal{H},\quad (\mathcal{U}\Psi \,\widetilde { \mid}\,\mathcal{U}\Phi )\equiv  (\widetilde { \Psi }  \ \widetilde { \mid} \ \widetilde { \Phi }  ) =(\Psi \mid \Phi ).
\ee
Under this unitary transformation, any quantum-mechanical operator such as, for example, the Hamiltonian operator $\mathrm{H}$, defined on $\mathcal{H}$ --- or rather on a dense subspace $\mathcal{D}$ of $\mathcal{H}$ --- is carried over to the pushforward operator $\breve{ \mathrm{H}}$ under $\mathcal{U}$, which is an operator defined on $\breve{ \mathcal{D}}\equiv \mathcal{U}(\mathcal{D})$:
\be\label{H breve}
\breve{ \mathrm{H}}\equiv \mathcal{U}\,\mathrm{H}\, \mathcal{U}^{-1}, 
\ee
that is from (\ref{psitilde=S^-1 psi}):
\be
\forall \Xi \in \breve{ \mathcal{D}}\equiv \mathcal{U}(\mathcal{D}), \quad \breve{ \mathrm{H}}\, \Xi = S^{-1} \mathrm{H} S\, \Xi .
\ee
The pushforward operator $\breve{ \mathrm{H}}$ {\it is physically equivalent to the starting operator} $\mathrm{H}$ since, from its definition (\ref{H breve}) and the unitarity of $\mathcal{U}$ (\ref{tilde=isometry}), all products $(\Psi \mid  \mathrm{H} \Phi  ), \Psi ,\Phi \in \mathcal{D}$, stay unchanged after the unitary transformation $\mathcal{U}$:
\be \label{H breve vs H}
\forall \Psi ,\Phi \in \mathcal{D},\quad (\mathcal{U}\Psi\ \widetilde{\mid}\ \breve{ \mathrm{H}} \, (\mathcal{U}\Phi) ) = (\Psi \mid \mathrm{H} \, \Phi ).
\ee
Note in particular that, for any state $\Psi \in \mathcal{D}$, the mean value of $ \mathrm{H}$ for this state, $\langle \mathrm{H} \rangle \equiv (\Psi \mid \mathrm{H} \, \Psi  )$, is equal to the mean value of $\breve{ \mathrm{H}}$ for the transformed state after the unitary transformation: $\langle \breve{ \mathrm{H}} \rangle \equiv (\mathcal{U}\Psi\ \widetilde{\mid}\ \breve{ \mathrm{H}} \, (\mathcal{U}\Psi ) )=\langle \mathrm{H} \rangle$. Because a sesquilinear form like $S(\Psi ,\Phi )\equiv (\mathcal{U}\Psi\ \widetilde{\mid}\ \breve{ \mathrm{H}} \, (\mathcal{U}\Phi) )$ is determined by the associated quadratic form,
\footnote{\
Indeed, let $\mathcal{S}(\Psi ,\Phi )$ be any sesquilinear form defined on some complex vector space $\mathcal{E}$, and let $Q(\Psi )\equiv \mathcal{S}(\Psi ,\Psi )\quad (\Psi \in \mathcal{E})$ be the associated quadratic form. Using merely the sesquilinearity of $\mathcal{S}$ ({\it i.e.} without assuming any symmetry property for $\mathcal{S}$), we get  for any $\Psi $ and $\Phi $ in $\mathcal{E}$:
\be\label{S from Q}
\mathcal{S}(\Psi , \Phi )=\{ iQ(\Psi +\Phi )+Q(\Psi +i\Phi )-(1+i)[Q(\Psi )+Q(\Phi )] \}/(2i).
\ee
}
this is {\it characteristic} of the pushforward operator: if an operator $\mathcal{O}$ defined on $\breve{ \mathcal{D}}\equiv \mathcal{U}(\mathcal{D})$ is such that 
\be \label{H breve vs O}
\forall \Psi \in \mathcal{D},\quad (\mathcal{U}\Psi\ \widetilde{\mid}\ \, \mathcal{O}(\mathcal{U}\Psi ) ) = (\Psi \mid \mathrm{H} \, \Psi  ),
\ee
then we have also (\ref{H breve vs H}) with $\mathcal{O}$ in the place of $\breve{ \mathrm{H}}$. As in Ref. \cite{A43}, Note 6, it follows then from the fact that $\mathcal{D}$ is a dense subspace of $\mathcal{H}$, that indeed $\mathcal{O}=\breve{ \mathrm{H}}$. \\

But the new operator, here the new Hamiltonian $\widetilde{\mathrm{H}}$, corresponding with the new coefficient field $\widetilde {\gamma}^\mu$ after the similarity transformation $S$, is defined in a different way than $\breve{ \mathrm{H}}$ --- although obviously it also acts on $\breve{ \mathcal{D}}\equiv \mathcal{U}(\mathcal{D})$. Namely, the new operator is defined as the starting one, but replacing the starting coefficient fields by the new ones. It seems natural to define that the two operators $\mathrm{H}$ and $\widetilde{\mathrm{H}}$ are physically equivalent iff they yield the same mean value for any two unitarily-equivalent states $\Psi $ and $\widetilde{\Psi }\equiv \mathcal{U}\Psi \equiv S^{-1}\Psi $. The following has thus been proved: {\it in order that the Hamiltonian operator $\widetilde{\mathrm{H}}$ after the similarity transformation $S$ be physically equivalent} in that sense {\it to the initial one $\mathrm{H}$, it is necessary and sufficient that we have $\widetilde{\mathrm{H}}=\breve{ \mathrm{H}}$, that is} \{\cite{A43}, Eq. (43)\}:
\be \label{similarity-invariance-H}
\widetilde{\mathrm{H} }  =  S^{-1}\,\mathrm{H}\, S.
\ee
It is clear from the derivation of this result that it applies exactly in the same way to any other quantum-mechanical operator, such as e.g. the energy operator defined in Eq. (\ref{E:=H^s}) below. Appendix \ref{ConstantShift} proves
 that, for the standard form of the covariant Dirac equation (DFW), the mean values of the energy operator cannot be shifted by a non-zero constant after an admissible similarity transformation. In other words, for DFW, the result proved above holds true if one accounts for a less restrictive \hyperref[Definition]{definition} of physically equivalent energy operators, which allows that all mean values can be shifted by the same constant.

\section{Non-uniqueness of the Hamiltonian}

The Hamiltonian operator [in a given coordinate system $(x^\mu )$] is defined by rewriting the Dirac equation in Schr\"odinger form. Thus (setting $\hbar =1$ in this paper):
\be \label{Schrodinger-general}
i \frac{\partial \Psi }{\partial t}= \mathrm{H}\Psi \qquad (t\equiv \frac{x^0}{c}),
\ee
and
\be \label{Schrodinger-general-tilde}
i \frac{\partial \Xi  }{\partial t}= \widetilde{\mathrm{H}}\Xi ,
\ee
respectively before and after application of a local similarity transformation $S$. As is also well known \cite{BrillWheeler1957+Corr,ChapmanLeiter1976}, the DFW equation is covariant (in a topologically-simple spacetime) under those local similarity transformations that are admissible, i.e., those that are got by lifting a local Lorentz transformation applied to the tetrad field. This means that, for a such admissible similarity transformation $S$, Eqs. (\ref{Schrodinger-general}) and (\ref{Schrodinger-general-tilde}) are equivalent if one exchanges the wave functions $\Psi $ and $\Xi $ according to the unitary transformation (\ref{psitilde=S^-1 psi}):
\be\label{chi = U psi}
\Xi = \mathcal{U}\Psi \equiv S^{-1}\Psi.
\ee
Substituting thus $\Psi=S \Xi$ into (\ref{Schrodinger-general}) and identifying with (\ref{Schrodinger-general-tilde}), we get:
\be\label{Htilde}
\widetilde{\mathrm{H}}=S^{-1} \mathrm{H}S-i\,S^{-1}\partial_t S=\breve{\mathrm{H}} -i\,S^{-1}\partial_t S. 
\ee
Comparing with (\ref{similarity-invariance-H}), we recover in a simple way \cite{A47} the result \{\cite{A43}, Eq. (48)\} that: {\it For DFW, in order that the Hamiltonian operator $\widetilde{\mathrm{H}}$ after the similarity transformation $S$ be physically equivalent to the initial one $\mathrm{H}$, it is necessary and sufficient that the similarity $S$ be independent of the time} $t$:
\be\label{partial_t S=0}
\partial_t S=0.
\ee
(In Ref. \cite{A43}, this had been got by comparing the explicit expressions of $\widetilde{\mathrm{H}}$ and $\mathrm{H}$.) {\it This result} and a similar result that gives the condition in order that the energy operators before and after the similarity be equivalent, {\it not} the ``demonstration that the form of Dirac Hamiltonians depends on the choice of tetrads", was the basis for our statement of the generic non-uniqueness of the DFW Hamiltonian and energy operators. Indeed, nothing prevents one from changing the coefficients by a {\it time-dependent} similarity, which leads hence to inequivalent Hamiltonians \cite{A43}. 

\paragraph {In summary:}\label{SummaryNonUniqueness} a change of the tetrad field defines an admissible similarity transformation $S$ that applies both to the field of Dirac matrices by (\ref{similarity-gamma}), and to the wave function by (\ref{psitilde=S^-1 psi}). The transformation (\ref{psitilde=S^-1 psi}) is in fact a unitary transformation $\,\mathcal{U}\,$ between the Hilbert spaces in which the wave function lives before and after the similarity transformation $S$, Eq. (\ref{tilde=isometry}). However, when $S$ depends on the time coordinate $t$, the new Hamiltonian operator $\widetilde{\mathrm{H}}$, after the similarity, does not coincide with the pushforward operator $\breve{\mathrm{H}}$ of the initial Hamiltonian H under the unitary transformation $\,\mathcal{U}$, Eq. (\ref{Htilde}). Hence, when $S$ depends on $t$, the mean values of H and $\widetilde{\mathrm{H}}$ cannot coincide for all states, so that H and $\widetilde{\mathrm{H}}$ are not physically equivalent. \\

The origin of the non-uniqueness problem is thus not trivial. It does not reside in the mere fact that many different fields of Dirac matrices can indifferently be chosen, nor in the other obvious fact that these different fields lead in general to different forms of the Hamiltonian operator in a given coordinate system. As I just showed anew, the non-uniqueness problem applies to the Hamiltonian associated with the standard version of the covariant Dirac equation, ``DFW" (as well as to the Hamiltonian associated with alternative versions of the covariant Dirac equation \cite{A43}), although the DFW equation itself has been carefully built so as to be covariant under the admissible similarity transformations, thus essentially unique. Only the latter uniqueness (the covariance of the DFW equation under the admissible similarity transformations) was included in the ``conclusions of previous studies \cite{BrillWheeler1957+Corr, ChapmanLeiter1976} on the independence of physical characteristics of the Dirac theory on the choice of tetrads", which Gorbatenko \& Neznamov \cite{GorbatenkoNeznamov2013} say to share. The non-uniqueness applies also \cite{A43} to the {\it energy operator} E, that coincides with the Hermitian part, for the scalar product (\ref{Hermitian-sigma=1-g}), of the Hamiltonian operator:
\be\label{E:=H^s}
\mathrm{E} = \frac{1}{2} (\mathrm{H}+\mathrm{H}^\ddagger ).
\ee
The non-uniqueness applies also to the {\it spectrum}  of the energy operator in a given coordinate system, as was proved in Ref. \cite{A43}, and this is also true in the presence of an electromagnetic field \cite{A48}.

\section{Trying to solve the non-uniqueness problem}

In a general coordinate system in a general spacetime, the metric depends on the time coordinate $t$, and then so does the field of orthonormal tetrads defining the field of Dirac matrices. It is then the general case that the local Lorentz transformation $L$ relating two different tetrad fields depend on $t$, so that the similarity transformation $S$ got by ``lifting" $L$ also depend on $t$, thus leading to a Hamiltonian $\widetilde{\mathrm{H}}$ that is not equivalent to the starting one H, see Eq. (\ref{Htilde}). This does not mean, of course, that {\it all} pairs $(\mathrm{H},\widetilde{\mathrm{H}})$, got by choosing one admissible tetrad field and transforming it to a new tetrad field through a local Lorentz transformation, are made of two inequivalent operators. {\it Therefore, exhibiting some pairs $(\mathrm{H}, \widetilde{\mathrm{H}})$ that are (supposedly) made of two equivalent operators, as do Gorbatenko \& Neznamov \cite{GorbatenkoNeznamov2013}, can not disprove the existence of the non-uniqueness problem.} In particular, in a coordinate system in which the metric is stationary: $g_{\mu \nu,0 }=0$, it is natural (though, of course, not mandatory) to choose time-independent tetrads, thus leading to coefficient fields related two by two by a time-independent similarity transformation, hence giving equivalent Hamiltonian operators. Such is the case for several among the examples given in Ref. \cite{GorbatenkoNeznamov2013}: most certainly for Example 4 based on the static, diagonal, space-isotropic metric considered among others by Obukhov \cite{Obukhov2001} and by Silenko \& Teryaev \cite{SilenkoTeryaev2005}, but likely also for some of the examples based on the Kerr metric, which is stationary.\\

Once the generic non-uniqueness of the Dirac Hamiltonian and energy operators is recognized, to escape this non-uniqueness demands to build some {\it prescription} that restrict the choice of tetrad when the coordinate system and the corresponding expression of the metric are given. That prescription should be consistent (i.e., well defined), moreover the corresponding restriction in the choice of the tetrad field should be sufficient (i.e., it should solve the problem), and preferably it should be physically motivated. Building a such prescription necessarily involves some choice, which is not strongly constrained in the current state of experimental knowledge. (Nevertheless, choosing tetrad fields with high rotation rates, without any relation to the rotation of a physical body, would lead to high theory-experiment discrepancies. Also recall that the non-uniqueness problem is already there in a flat spacetime as soon as one uses the DFW equation with its gauge freedom, and this also in the presence of the electromagnetic field, so that even the energy levels of the hydrogen atom would not be defined  \cite{A48}.) However, just finding one definite prescription (even an artificial, computationally-motivated one) that {\it really} provide unique Hamiltonian and energy operators, is difficult. \\

In a series of papers by Gorbatenko \& Neznamov \cite{GorbatenkoNeznamov2010, GorbatenkoNeznamov2011, GorbatenkoNeznamov2011b}, attempts have been made at finding a such prescription. 
\footnote{\ 
The very title of the paper \cite{GorbatenkoNeznamov2010} indicates clearly that its authors aimed at solving some non-uniqueness problem for the Dirac Hamiltonian in a curved spacetime. To the best of my knowledge, the existence of a such problem has been noted for the first time in Ref. \cite{A42}, and shown in detail in Ref. \cite{A43}.
} 
Their first paper \cite{GorbatenkoNeznamov2010} was limited to a time-independent metric. For that case, as recalled above, one might content oneself with choosing time-independent tetrad fields, since any two of them lead to equivalent Hamiltonian and to equivalent energy operators \cite{A43}. Their proposal for the general, time-dependent case \cite{ GorbatenkoNeznamov2011, GorbatenkoNeznamov2011b} was discussed in detail in Ref. \cite{A47}. As noted there, that proposal consists first in going from the starting arbitrary [orthonormal] tetrad field, say $(u_\alpha )$, to what the authors name ``the Schwinger tetrad", and which is a tetrad field, say  $(\widetilde{u}_\alpha )$, of which each among the three ``spatial" vector (fields) $\widetilde{u}_p \ (p=1,2,3)$ has a zero ``time" component, in the coordinate system considered. Their procedure based on the formalism of pseudo-Hermitian Hamiltonians leads them then to define as the candidate for a unique Hamiltonian the one noted $\mathrm{H}_\eta$, got after a similarity transformation denoted $\eta $ by them, and which in the form (\ref{similarity-gamma}) I will denote by $T=\eta ^{-1}$. We have \{Ref. \cite{GorbatenkoNeznamov2011}, Eq. (66)\}:
\be\label{T or eta^-1}
T =a^{-1} S \equiv (\eta _\mathrm{G\&N})^{-1},
\ee
where $S$ is the admissible similarity transformation associated with the change of tetrad from $(u_\alpha )$ to  $(\widetilde{u}_\alpha )$, and $a=\abs{g\,g^{00}}^{1/4}$. ($S^{-1}$ is what these authors note $L$.) It is easy to see, as they also note \cite{GorbatenkoNeznamov2011b}, that the Hamiltonian $\mathrm{H}_\eta$ got after the similarity transformation $T$ is {\it also} equal to the {\it energy operator} [Eq. (\ref{E:=H^s}) in the present paper] got with the field of Dirac matrices deduced from the starting field by the admissible similarity transformation $S$. \hypertarget{G&N-Prescription}{Note that their procedure} is really a {\it prescription} for restricting the choice of tetrad field in order to (try to) get unique Hamiltonian and energy operators, of the general kind I described in the foregoing paragraph. Indeed, not every tetrad field is a ``Schwinger tetrad" in a given coordinate system. Thus, {\it if their procedure would actually lead to a unique Hermitian Hamiltonian in a given coordinate system, that would not prove anything against the existence of the non-uniqueness problem:} that would be a first check of a prescription for solving it.
\footnote{\
As shown in Ref. \cite{A48}, Sect. 4, it is actually not enough to get unique Hamiltonian and energy operators in any given coordinate system, for what is  physically given is the reference frame (a three-dimensional congruence of time-like world lines), not the coordinate system, for which there is a vast functional space of different choices within a given reference frame.
}
But that is not the case, as I will now show.

\section{The ``Schwinger tetrad" prescription is not unique}\label{Schwinger not unique}

As I showed in detail in Ref. \cite{A47}, App. C, the choice of a tetrad that, in a given coordinate system, is a ``Schwinger tetrad", is far from unique. Gorbatenko \& Neznamov did not prove that, if in the same coordinate system one takes a second Schwinger tetrad, then the Hermitian Hamiltonian provided by their construction from that second Schwinger tetrad is physically equivalent to the Hermitian Hamiltonian got from the first Schwinger tetrad, {\it and indeed that is not generally the case.} In Ref. \cite{A47}, App. C, I made remarks that indicated this fact, and I gave a precise counterexample to illustrate this. Now I will prove a general result. \\

After the similarity transformation (\ref{T or eta^-1}), the scalar product becomes the ``flat" one, i.e., $\sqrt{-g}\, \widetilde{A}\, \widetilde{\gamma}^0={\bf 1} _4$ in Eq. (\ref{Hermitian-tilde}), as stated by Gorbatenko \& Neznamov. This results from Eq. (65) in Ref. \cite{GorbatenkoNeznamov2011}, and from the fact that, after a general similarity transformation $T$, the matrix $M\equiv \sqrt{-g} A \gamma ^0$ transforms like this [see Eqs. (\ref{similarity-gamma}) and
(\ref{Hermitian-tilde})$_2$]:
\be \label{similarity-M}
\widetilde{M}\equiv \sqrt{-g}\,  \widetilde{A}\, \widetilde{\gamma}^0 =T^\dagger  M T.
\ee
Now, besides the first ``Schwinger tetrad" $(\widetilde{u}_\alpha )$, consider another one, $(\check{u}_\alpha )$, and let $S'$ be the admissible similarity transformation associated with the change of tetrad from the starting arbitrary tetrad $(u_\alpha )$ to $(\check{u}_\alpha )$. As in Eq. (\ref{T or eta^-1}), let $T'=a^{-1}S'=\eta'^{-1}$ be the similarity transformation leading to the other candidate Hamiltonian $\mathrm{H}_{\eta'}$, corresponding to this other choice of a Schwinger tetrad. The change from the Hamiltonian $\mathrm{H}_\eta$ to the Hamiltonian $\mathrm{H}_{\eta'}$ is through the similarity 
\be
U=T'\,T^{-1}=S'\,S^{-1},
\ee
which is admissible as are $S$ and $S'$. Since the matrix $M$ is equal to ${\bf 1}_4$ before and after the application of $U$, it results from the transformation law (\ref{similarity-M}) that $U=U(X)$ is a unitary {\it matrix,} $U^\dagger\, U={\bf 1}_4$. The operators $\mathrm{H}_\eta$ and $\mathrm{H}_{\eta'}$ can be said physically equivalent iff Eq. (\ref{similarity-invariance-H}) is verified, which writes here:
\be \label{similarity-invariance-Htilde}
\mathrm{H}_{\eta'}  =  U^{-1}\,\mathrm{H}_\eta\, U.
\ee
Let us determine when this is true. Since $\mathrm{H}_\eta$ (respectively $\mathrm{H}_{\eta'}$) is equal to the energy operator got with the tetrad $(\widetilde{u}_\alpha )$ (respectively with the tetrad $(\check{u}_\alpha )$), these two operators exchange by the admissible similarity $S'\,S^{-1}=U$. The condition in order that two DFW energy operators related by an admissible similarity transformation $U$ be equivalent, Eq. (\ref{similarity-invariance-Htilde}), is given by Eq. (64) in Ref. \cite{A43}:\be\label{SEtilde=ES DFW}
B(\partial _t U)U^{-1}-\left[B(\partial _t U)U^{-1}\right]^\dagger = 0, \qquad B\equiv A\gamma ^0,
\ee 
with $\gamma ^\mu $ the field of Dirac matrices for the first energy operator. When the first tetrad, corresponding with that field $\gamma ^\mu $, is a ``Schwinger tetrad", we have $B= \sqrt{g^{00}}\, {\bf 1}_4$ \{Ref. \cite{A43}, Eq. (78). We assume here that $A=\gamma ^{\natural 0}$, as is standard: see Note \ref{A constant}.\} Thus the condition in order that we have (\ref{similarity-invariance-Htilde}) is simply:
\be
(\partial _t U)U^{-1}=\left[(\partial _t U)U^{-1}\right]^\dagger.
\ee
But since here $U$ is a unitary matrix, it is immediate to check that this is true if and only if $\partial _t U =0$. We have proved the following: \\

\noi (\hypertarget{Point i}{{\bf i})} {\it The energy operators $\,\mathrm{H}_\eta\,$ and $\,\mathrm{H}_{\eta'}\,$ corresponding with two different tetrads, each of which is a Schwinger tetrad in the same given coordinate system, exchange by an admissible similarity transformation whose matrix $\,U(X)\,$ is unitary.} \\

\noi (\hypertarget{Point ii}{{\bf ii})} {\it In order that $\,\mathrm{H}_\eta\,$ and $\,\mathrm{H}_{\eta'}\,$ be physically equivalent, it is necessary and sufficient that $\,\partial _t U =0$.}\\

\noi However, once again, it turns out to be the general case that $U$ depend on $t\equiv x^0/c$. Indeed, given that $(\widetilde{u}_\alpha )$ is a Schwinger tetrad in some coordinate system $(x^\mu )$, and given a local Lorentz transformation $L$, the necessary and sufficient condition in order that the tetrad $\check{u}_\beta = L^\alpha _{\ \, \beta }\,\widetilde{u}_\alpha$ be also a Schwinger tetrad in the same coordinate system is that \{Ref. \cite{A47}, Eq. (89)\}:
\be\label{L^0_p=0}
L^0 _{\ \, p }=0 \quad (p=1,2,3).
\ee
In general, a local Lorentz transformation $L$ verifying this condition depends on $t$, and so does the associated admissible similarity transformation $U$, got from $L$ by using the spinor representation ${\sf S}$ (defined up to a sign): $U=\pm {\sf S}(L)$.

\section{The spin-rotation coupling}

In their Examples 6 and 7, Gorbatenko \& Neznamov \cite{GorbatenkoNeznamov2013} comment on my discussion (Ref. \cite{A49}, Sect. 4) of the DFW Hamiltonians got in two reference frames in a Minkowski spacetime, when using three different tetrad fields. 
Their comment is limited to two tetrad fields: i) $u'_\alpha \equiv \delta ^\mu _\alpha \partial '_\mu $, that is, the natural basis $(\partial '_\mu )$ of a Cartesian coordinate system $(x'^\mu )=(ct',x',y',z')$, or ``Cartesian tetrad"; and ii) the tetrad $(u_\alpha)$ got from $(u'_\alpha )$ by a spatial rotation of angle $\omega t$ around the axis $Oz'$, with $\omega $ a real constant \{Ref. \cite{A49}, Eqs. (34)--(35)\}. Using Eqs. (33) and (35) of Ref. \cite{A49}, one checks immediately that both tetrads $(u'_\alpha)$ and $(u_\alpha )$ are Schwinger tetrads in the Cartesian coordinates, as well as in the rotating coordinates $(x^\mu)=(ct,x,y,z)$ given by
\be\label{rotating Cartesian}
t=t',\quad x=x'\cos \omega t + y' \sin \omega t,\quad y=-x' \sin \omega t + y' \cos \omega t,\quad z=z'.
\ee 
Therefore, the discussion in Sect. \ref{Schwinger not unique} applies. Moreover, as I noted \cite{A49}, the Hamiltonians with the Cartesian tetrad $(u'_\alpha )$: $\mathrm{H}'_1$ in the inertial frame and $\mathrm{H}_1$ in the rotating frame, are Hermitian, as are also those with the rotating tetrad $(u_\alpha )$: $\mathrm{H}'_3$ in the inertial frame and $\mathrm{H}_3$ in the rotating frame. Thus each among these Hamiltonians coincides with the corresponding energy operator.\\

Example 6 in Ref. \cite{GorbatenkoNeznamov2013} comments on the Hamiltonians in the inertial reference frame, $\mathrm{H}'_1$ and $\mathrm{H}'_3$ \{respectively Eqs. (26) and (66) in Ref. \cite{A49}\}, which are rewritten by the authors of Ref. \cite{GorbatenkoNeznamov2013} as their Eqs. (28) and (30), respectively. As I noted already in Ref. \cite{A47} [Eq. (94) there], the two tetrads $(u'_\alpha)$ and $(u_\alpha )$ exchange by a time-dependent Lorentz transformation $L=L(t)$: namely, the rotation of angle $\omega t$ around the $Oz'$ axis. Hence, the corresponding Hamiltonians $\mathrm{H}'_1=H'_{\mathrm{G\&N}}$ and $\mathrm{H}'_3=H_{\mathrm{G\&N}}$ exchange by the time-dependent similarity transformation $S(t)=\pm {\sf S}(L(t))$, {\it and thus are not physically equivalent \cite{A47}, contrary to what Gorbatenko \& Neznamov \cite{GorbatenkoNeznamov2013} state}. This is just confirmed by their Eqs. (32) and (33): the similarity matrix $S(t)$ is what these authors note $R^{-1}=R^\dagger $, it is indeed a unitary {\it matrix} as proved generally at \hyperlink{Point i}{Point (i)} in Sect. \ref{Schwinger not unique} --- but as proved at \hyperlink{Point ii}{Point (ii)} the energy operators $\mathrm{H}_\eta \equiv \mathrm{H}'_1=H'_{\mathrm{G\&N}}$ and $\mathrm{H}_{\eta'} \equiv \mathrm{H}'_3=H_{\mathrm{G\&N}}$ are {\it not} physically equivalent. Note that their Eq. (33):
\be\label{33 G&N}
H=RH'R^\dagger -iR\frac{\partial R^\dagger }{\partial t}
\ee
is a particular case of Eq. (\ref{Htilde}) above, corresponding with $S=R^\dagger =R^{-1}$. This is of course expected, because $H'$ and $H$ exchange by the admissible similarity transformation $S\equiv R^\dagger,\ S^{-1}=R$. Therefore, since $S$ does not verify the condition (\ref{partial_t S=0}) for physically equivalent operators, we know \hyperlink{Mean depends}{from the end of Appendix A} that the energy mean values got from $H$ and from $H'$ differ by a number which {\it depends} on the state. It is interesting to check this explicitly.\\

The operator $RH'R^\dagger$ on the r.h.s. of (\ref{33 G&N}) is the pushforward operator of the operator $H'$ under the unitary transformation $\ \mathcal{U}$ associated with the similarity transformation $S=R^\dagger$. Hence its mean value for the transformed state $\psi=R\psi '$ equals the mean value of $H'$ for the starting state $\psi '$,  for any state $\psi '$. From the explicit expression $R=e^{\omega tN}$ with $N$ a constant matrix: $N\equiv (\alpha '^1\alpha '^2)/2\ $ \cite{GorbatenkoNeznamov2013} (with $N^\dagger=-N$), it follows for the additional term: $R\frac{\partial R^\dagger }{\partial t}=-e^{\omega tN}\omega Ne^{-\omega tN}=-\omega N$. Therefore:
\be\label{bar A}
\langle H \rangle - \langle H' \rangle =\int \psi ^\dagger \left(i\omega N \right)\psi \,\dd^3{\bf x}.
\ee
According to the notation of Ref. \cite{GorbatenkoNeznamov2013}, $\alpha '^j\ (j=1,2,3)$ are the Dirac ``alpha" matrices for a flat Minkowski spacetime, thus $\alpha '^j=\gamma '^0\gamma '^j\ $ with $\gamma '^\mu  \ (\mu =0,...,3)$ the Dirac ``gamma" matrices for a flat Minkowski spacetime. \{Cf. Ref. \cite{A49}, Eq. (26).\} Thus $N\equiv \frac{1}{2}\alpha '^1\alpha '^2=-\frac{1}{2}\gamma'^1 \gamma'^2$. Assuming the standard Dirac ``gamma" matrices are chosen, we get then: $
N=\frac{i }{2}\Sigma ^3=\frac{i }{2}\mathrm{diag}(1,-1,1,-1)$, 
whence from (\ref{bar A}):
\be
\langle H \rangle - \langle H' \rangle =-\frac{\omega}{2}\int \psi ^\dagger\, \Sigma ^3\, \psi \,\dd^3{\bf x} \equiv -\frac{\omega}{2}\langle \Sigma ^3 \rangle.
\ee
\hypertarget{H'_1 not equiv H'_3}{This gives with} $\psi =(\psi ^\alpha )_{\alpha =0,...,3}\,$:
\be\label{bar A-explicit}
A\equiv \langle H \rangle - \langle H' \rangle =-\frac{\omega}{2}\int \left(\abs{\psi^0}^2 +\abs{\psi^2}^2-\abs{\psi^1}^2-\abs{\psi^3}^2\right)\,\dd^3{\bf x}.
\ee
Thus, the difference in the mean values of the energy operators $H$ and $H'$ for corresponding states $\psi '$ and $\psi =R\psi '$ {\it depends on the state $\psi $} and contains the {\it arbitrary} factor $\omega $. (Indeed, when looking for energy mean values in the inertial frame, the rotation velocity of the tetrad $(u_\alpha )$ is absolutely arbitrary.) 
For a normed state: $(\psi \mid \psi )=\int \left(\abs{\psi^0}^2 +\abs{\psi^1}^2+\abs{\psi^2}^2+\abs{\psi^3}^2\right)\,\dd^3{\bf x}=1$, 
$A$ can take any value between $-\frac{\omega}{2}$ and $+\frac{\omega}{2}$. This means simply that we cannot predict anything about mean energy. So the two energy operators $H$ and $H'$ are grossly inequivalent from the physical point of view.\\ 

Another way of exhibiting the physical inequivalence of $H'$ and $H$ is by looking at their energy eigenstates. Suppose $\psi '$ is an eigenstate for $H'$, with eigenvalue $e'$: $H'\psi '=e'\psi '$. We get for the corresponding state $\psi \equiv R\psi '$: 
\be
RH'R^\dagger (R\psi ')=RH'\psi ' =e'(R\psi ').
\ee
Since 
\be H=RH'R^\dagger -iR\frac{\partial R^\dagger }{\partial t}=RH'R^\dagger+i\omega N=RH'R^\dagger -\frac{\omega  }{2}\Sigma ^3, 
\ee
we therefore get for the corresponding state $\psi \equiv R\psi '$:
\be
H\psi =e'\psi -\frac{\omega  }{2}\Sigma ^3 \psi .
\ee
But $\Sigma ^3 \psi\ne 0$ unless $\psi =0$. Thus, the corresponding state $\psi \equiv R\psi '$ is {\it not} an eigenstate for the same eigenvalue $e'$. In fact, even for a given eigenstate $\psi' $ of $H'$, the difference $H\psi -e'\psi$ is again {\it arbitrarily large}.\\

Example 7 in Ref. \cite{GorbatenkoNeznamov2013} comments on the Hamiltonians in the uniformly rotating reference frame, $\mathrm{H}_1$ and $\mathrm{H}_3$ \{respectively Eqs. (32) and (70) in Ref. \cite{A49}\}, which are rewritten by the authors of Ref. \cite{GorbatenkoNeznamov2013} as their Eqs. (38) and (40), respectively. Nearly the same can be written as for Example 6, because the two tetrads and hence the similarity $S$ are the same in the two examples, and, since $S=S(t)$, we have $\partial_t S=\partial _{t'} S$ from Eq. (\ref{rotating Cartesian}) above. A difference is that now $\omega $ is fixed for a given physical situation, since through Eq. (\ref{rotating Cartesian}) it defines the very rotating frame, in which we are trying to calculate the energy mean values. However, we might provide the tetrad $(u_\alpha )$ with a different rotation velocity, say $\omega '$, as compared with that of the rotating frame. Thus again $\omega '$ would be arbitrary. \\

Now we have the fact \cite{A49} that the spin-rotation coupling term $-\frac{\hbar \omega }{2}\Sigma ^3=-\Mat{\omega }.{\bf S }$ \cite{Mashhoon1988,HehlNi1990} is indeed involved in one among two Hamiltonians/energy operators in the uniformly rotating reference frame: $\mathrm{H}_3$, but not in the other one: $\mathrm{H}_1$ (see also Ryder \cite{Ryder2008}). This fact cannot be discarded by stating that ``the spin-rotation coupling has no effect on the final physical characteristics of the quantum mechanical systems under consideration" \cite{GorbatenkoNeznamov2013}, because the two energy operators $\mathrm{H}_3$ and $\mathrm{H}_1$ are not physically equivalent. Nor, for the same reason, can the surprising fact \cite{A49} that the Hamiltonian/energy operator $\mathrm{H}'_3$ in the {\it inertial frame} does have the spin-rotation coupling term be discarded. But these two facts should lead one to ask whether this term must be there or not. In the Conclusion of Ref. \cite{A49}, I explain why I believe that the answer has to be experimental: it amounts to empirically deciding between two non-equivalent prescriptions \cite{A47,A48} for solving the non-uniqueness problem.

 \section{Summary}
 
The reason for the non-uniqueness problem \cite{A43} has been reexplained and \hyperref[SummaryNonUniqueness]{summarized} by appealing precisely to the notions of a unitary transformation and of the mean value of an operator, invoked by Gorbatenko \& Neznamov \cite{GorbatenkoNeznamov2013}. Their arguments actually aim at proving, not the uniqueness of the covariant Dirac theory, but the uniqueness that would be got (in their opinion) by using their particular \hyperlink{G&N-Prescription}{prescription} \cite{GorbatenkoNeznamov2011, GorbatenkoNeznamov2011b} to select the tetrad field. Although I showed this already by exhibiting a counterexample \cite{A47}, I showed here in a more general way that \hyperlink{Point i}{their prescription does not solve} the non-uniqueness problem. Finally, the non-uniqueness of the Hamiltonian cannot be disproved by exhibiting some pairs of equivalent Hamiltonians. However, the examples regarding my discussion \cite{A49} of the spin-rotation coupling are made of pairs of \hyperlink{H'_1 not equiv H'_3}{grossly non-equivalent} Hamiltonians.\\

\noi {\bf Acknowledgement.} It was noted by M. V. Gorbatenko \& V. P. Neznamov (private communication) and by a referee that, in the first version of this paper, it was not accounted for the fact that the energy can usually be subjected to a constant shift. The referee suggested a definition of physically equivalent energy operators which is equivalent to the one given below.

\appendix
\section{Appendix: Can the energy be shifted by a constant?}\label{ConstantShift}

It is generally considered that the energy of a quantum-mechanical system is defined only up to a real constant. Indeed, if we replace the wave function $\Psi $ by 
\be\label{Psi tilde}
\widetilde{\Psi}(t,x)\equiv e^{-iCt}\, \Psi (t,x),
\ee
with $C$ a real constant, then the starting Schr\"odinger equation (\ref{Schrodinger-general}) is equivalent to the following one:
\be \label{Schrodinger-general-tilde-2}
i \frac{\partial \widetilde{\Psi} }{\partial t}= \widetilde{\mathrm{H}}\widetilde{\Psi},
\ee
with the new Hamiltonian
\be \label{H-tilde}
\widetilde{\mathrm{H}}\equiv \mathrm{H}+C.
\ee
Then, also the energy operator E [Eq. (\ref{E:=H^s})] is replaced by $\widetilde{\mathrm{E}}\equiv \mathrm{E}+C$. Thus, all energy eigenvalues and mean values are just shifted by one and the same constant $C$. Note that the foregoing applies to any quantum-mechanical Hamiltonian. Coming back to the covariant Dirac equation, the transformation (\ref{Psi tilde}) can be seen as a local similarity transformation, for which the matrix $S$ in Eqs. (\ref{similarity-gamma}) and (\ref{psitilde=S^-1 psi}) has the special form
\be\label{T}
T=e^{iCt}\,{\bf 1}_4.
\ee
[This leaves the $\gamma ^\mu $ matrices unchanged by Eq. (\ref{similarity-gamma}).] Note that indeed the transformation law (\ref{Htilde}) of the Hamiltonian after a similarity transformation gives again (\ref{H-tilde}) when it is applied to the particular transformation (\ref{T}). 
\footnote{\
This occurs because $\widetilde{\mathrm{H}}$ in Eq. (\ref{Schrodinger-general-tilde-2}) is defined so that Eqs. (\ref{Schrodinger-general}) and (\ref{Schrodinger-general-tilde}) are equivalent. But $\mathrm{tr}\, \frac{\partial  T}{\partial  t}T^{-1}=4iC$, hence $T$ is not an admissible similarity transformation for DFW if $C \ne 0$: see Eqs. (\ref{dS/dt S^-1}) and (\ref{tr s alpha beta = 0}).
}
\\

Recall that it is the energy operator E which is relevant to the energy, and that for the covariant Dirac equation it is in general different from the Hamiltonian H. Therefore, we might make the definition of equivalent energy operators less restrictive by adopting the following one:

\paragraph {\bf Definition.}\label{Definition} {\it The energy operators $\mathrm{E}$ and $\widetilde{\mathrm{E}}$, before and after the unitary transformation $\,\mathcal{U}$ associated with a similarity transformation $S$ [Eq. (\ref{psitilde=S^-1 psi})], are physically equivalent iff there is a real constant $C$ such that the energy mean values are just shifted by $C$}:
\be \label{E breve vs E-C-norm 1}
\forall \Psi \in \mathcal{D} \ \mathrm{with}\ (\Psi \mid \Psi  )=1,\quad (\mathcal{U}\Psi\ \widetilde{\mid}\ \, \widetilde{\mathrm{E}}(\mathcal{U}\Psi ) ) - (\Psi \mid \mathrm{E} \, \Psi  )=C.
\ee

\vspace{3mm}
\noi By extending and formalizing the line of reasoning used around Eqs. (\ref{H breve vs H})--(\ref{H breve vs O}), we shall first prove the following:

\paragraph{Lemma.}\label{Lemma} {\it In order that we have (\ref{E breve vs E-C-norm 1}), it is necessary and sufficient that}
\be\label{Etilde=Ebreve+C}
\widetilde{\mathrm{E}}=S^{-1}\,\mathrm{E}\,S +C{\bf 1}_4 .
\ee

\vspace{3mm}
\noi {\it Proof.} Consider the sesquilinear forms defined on $\mathcal{D}\times \mathcal{D}$:
\be
\mathcal{S}(\Psi ,\Phi )\equiv (\mathcal{U}\Psi\ \widetilde{\mid}\ \, \widetilde{\mathrm{E}}(\mathcal{U}\Phi ) ) 
\ee
and
\be
\mathcal{S}'(\Psi ,\Phi )\equiv (\Psi \mid \mathrm{E}\Phi  ) +C (\Psi \mid \Phi  ).
\ee

\vspace{3mm}
\noi Denoting $Q(\Psi )\equiv \mathcal{S}(\Psi ,\Psi  )$ and $Q'(\Psi )\equiv \mathcal{S}'(\Psi ,\Psi  )$ the associated quadratic forms defined on $\mathcal{D}$, (\ref{E breve vs E-C-norm 1}) is equivalent to:
\be \label{Q=Q' on sphere}
\forall \Psi \in \mathcal{D} \ \mathrm{with}\ (\Psi \mid \Psi  )=1,\quad Q(\Psi )=Q'(\Psi ).
\ee
Due to the homogeneity of degree 2, this of course means that $Q=Q'$. Hence, by (\ref{S from Q}), this is equivalent to $\mathcal{S}=\mathcal{S}'$. Thus, (\ref{E breve vs E-C-norm 1}) is equivalent to:
\be\label{S=S'}
\forall \Psi,\Phi  \in \mathcal{D},\quad (\mathcal{U}\Psi\ \widetilde{\mid}\ \, \widetilde{\mathrm{E}}(\mathcal{U}\Phi ) ) = (\Psi \mid \mathrm{E}\Phi  ) +C (\Psi \mid \Phi  ).
\ee
As with (\ref{H breve vs H}), the pushforward energy operator $\breve{ \mathrm{E}}=S^{-1}\mathrm{E}S$ verifies:
\be \label{E breve vs E}
\forall \Psi ,\Phi \in \mathcal{D},\quad (\mathcal{U}\Psi\ \widetilde{\mid}\ \breve{ \mathrm{E}} \, (\mathcal{U}\Phi) ) = (\Psi \mid \mathrm{E} \, \Phi ).
\ee
Hence, using also the unitarity (\ref{tilde=isometry}) of $\mathcal{U}$, (\ref{S=S'}) may be rewritten as:
\be\label{S=S'-2}
\forall \Psi,\Phi  \in \mathcal{D},\quad (\mathcal{U}\Psi\ \widetilde{\mid}\ \, \widetilde{\mathrm{E}}(\mathcal{U}\Phi ) ) = (\mathcal{U}\Psi\ \widetilde{\mid}\  \breve{ \mathrm{E}}(\mathcal{U}\Phi  )) +C (\mathcal{U}\Psi\ \widetilde{\mid}\  \mathcal{U}\Phi  ).
\ee
Because $\breve{ \mathcal{D}}\equiv \mathcal{U}(\mathcal{D})$, which is the domain of each among the two operators $\widetilde{\mathrm{E}}$ and $\breve{ \mathrm{E}}$, is in one-to-one correspondence with $\mathcal{D}$ under $\mathcal{U}$, this is still equivalent to
\be\label{S=S'-3}
\forall \Xi ,\Omega   \in \breve{ \mathcal{D}},\quad (\Xi \ \widetilde{\mid}\ \, \widetilde{\mathrm{E}}\Omega  ) = (\Xi \ \widetilde{\mid}\  \breve{ \mathrm{E}}\Omega)  +C (\Xi \ \widetilde{\mid}\  \Omega   )\equiv (\Xi \ \widetilde{\mid}\  (\breve{ \mathrm{E}}+C{\bf 1}_4)\Omega).
\ee
But since $\breve{ \mathcal{D}}$ is dense in the Hilbert space $\widetilde{ \mathcal{H}}$, the latter means that we have $\widetilde{\mathrm{E}}=\breve{ \mathrm{E}}+C{\bf 1}_4$, that is, precisely (\ref{Etilde=Ebreve+C}). The Lemma is proved.\\

\vspace{3mm}
\noi However, it turns out that (\ref{Etilde=Ebreve+C}) can happen after an {\it admissible} local similarity transformation only if $C=0$:

\paragraph{Theorem.} \label{Theorem} {\it Suppose that the local similarity transformation $S$ is admissible, i.e., $S(X) \in {\sf Spin(1,3)}$ for any $X \in \mathrm{V}$. If the energy operators $\mathrm{E}$ and $\widetilde{\mathrm{E}}$ before and after the application of $S$ are physically equivalent in the sense of (\ref{E breve vs E-C-norm 1}), then $C=0$.}\\

\noi {\it Proof.} From the \hyperref[Lemma]{Lemma}, we know that the condition (\ref{E breve vs E-C-norm 1}) is equivalent to (\ref{Etilde=Ebreve+C}). Setting $\delta \mathrm{E}\equiv S\widetilde{\mathrm{E}}S^{-1} -\mathrm{E}$, Eq. (\ref{Etilde=Ebreve+C}) is equivalent to
\be\label{deltaE = C I}
\delta \mathrm{E} = C {\bf 1}_4.
\ee
In Ref. \cite{A43}, Eq. (73), the following expression has been derived most generally for $\delta \mathrm{E}$:
\be\label{SEtilde-ES-DFW-1}
\delta \mathrm{E} =-iB^{-1} \left[B(\partial _0S)S^{-1}\right]^a, \qquad Q^a\equiv \frac{1}{2}(Q-Q^\dagger ),
\ee
where $B\equiv A\gamma ^0$. Note that $S$ is a smooth function of $X$, as is by definition a local similarity transformation. Hence, the assumption that $S(X) \in {\sf Spin(1,3)}$ for all $X$ implies that $S$ is deduced from the smooth local Lorentz transformation $L\equiv \Lambda \circ S$. (Here $\Lambda :{\sf Spin(1,3)}\rightarrow {\sf SO(1,3)}$ is the two-to-one covering map of the special Lorentz group by the spin group.) In other words, $S$ is got from a smooth change of the tetrad field by using the spinor representation. Using the fact that the explicit form of the spinor representation (e.g. Ref. \cite{Schulten1999}) is generated by the commutators of the ``flat" Dirac matrices $\gamma ^{\natural \alpha} $:
\be\label{s^ab}
s^{\alpha \beta } \equiv [\gamma^{\natural \alpha}  ,\gamma^{\natural \beta }  ],
\ee
it has been shown in Ref. \cite{A43} [Eq. (81) and below there] that we have then
\be\label{dS/dt S^-1}
\frac{\partial  S}{\partial  t}S^{-1}= \omega_{\alpha \beta} s^{\alpha \beta },
\ee
in which the six coefficients $\omega_{\alpha \beta}=-\omega_{\beta \alpha }$ are real and depend on the spacetime point $X$ as does $S$. Hence, we may rewrite (\ref{SEtilde-ES-DFW-1}) more explicitly as:
\be\label{deltaE}
2\delta \mathrm{E} =-iB^{-1} \left[B(\omega_{\alpha \beta}s^{\alpha \beta })-(B \omega_{\alpha \beta}s^{\alpha \beta })^\dagger\right]=-i\omega_{\alpha \beta} \left[s^{\alpha \beta }-B^{-1}\left(s^{\alpha \beta }\right)^\dagger\,B^\dagger\right]. 
\ee
Now, from the definition (\ref{s^ab}), and since $\mathrm{tr}\,MN=\mathrm{tr}\,NM$ for two square matrices $M$ and $N$ having the same dimension, we have:
\be\label{tr s alpha beta = 0}
\mathrm{tr}\,s^{\alpha \beta }=0,
\ee
hence also $\mathrm{tr}\,(s^{\alpha \beta })^\dagger=0$. We have also $B^\dagger=B$ from the definition of the hermitizing matrix $A$ \{Ref. \cite{A43}, Eq. (7)\}. Therefore, it follows from Eq. (\ref{deltaE}) that
\be\label{tr deltaE = 0}
\mathrm{tr}\, \delta \mathrm{E}=0.
\ee
However, if Eq. (\ref{deltaE = C I}) is verified, we have $\mathrm{tr}\, \delta \mathrm{E}=4C$, hence $C=0$ from (\ref{tr deltaE = 0}). This proves the Theorem.\\

\hypertarget{Mean depends}{That} \hyperref[Theorem]{Theorem} says that, for the standard version of the covariant Dirac equation (DFW), the energy mean values can {\it not} be shifted by a constant number. If a local similarity tranformation is admissible for DFW and leaves the energy mean values unchanged up to a constant shift as in Eq. (\ref{E breve vs E-C-norm 1}), then the new energy operator $\widetilde{\mathrm{E}}$ is just the pushforward operator  $\breve{ \mathrm{E}}$ that leaves the energy mean values exactly unchanged. Said differently: if we have $\widetilde{\mathrm{E}} \ne \breve{ \mathrm{E}}$, then the mean values of the energy operators $\mathrm{E}$ and  $\widetilde{\mathrm{E}}$ before and after the similarity transformation differ by a number which {\it depends} on the state $\Psi $. This result is true also for the Hamiltonian operator. (The proof is almost the same but a bit simpler.)


\end{document}